\documentclass[aps,prb,twocolumn,superscriptaddress]{revtex4-1}

\usepackage{graphicx}
\usepackage{hyperref}
\usepackage{amsfonts}
\usepackage{amsmath,bm,amssymb}
\usepackage{mathtools}
\usepackage{comment}

\begin{document}
\DeclarePairedDelimiter\bra{\langle}{\rvert}
\DeclarePairedDelimiter\ket{\lvert}{\rangle}
\DeclarePairedDelimiterX\braket[2]{\langle}{\rangle}{#1 \delimsize\vert #2}

%\definechangesauthor[color=red]{YC}
%\definechangesauthor[color=blue]{JV}
%\definechangesauthor[color=green]{KP}
%\added[id=<id>,remark=<remark>]{text}
%\deleted[id=<id>,remark=<remark>]{text}
%\replaced[id=<id>,remark=<remark>]{newtext}{oldtext}
%\listofchanges[style=<list|summary>]
%remove markup: \usepackage[final]{changes}
%ctan.net has script to remove markup scripts

\title{Zeeman-splitting-induced Topological Nodal Structure and Anomalous Hall
Conductivity in ZrTe$_5$}

\author{Yichul Choi}
\email{yichul00@vt.edu}
\affiliation{Department of Physics, Virginia Tech, Blacksburg, Virginia 24061, USA}
\author{John W. Villanova}
\email{jvillano@uark.edu}
\affiliation{Department of Physics, University of Arkansas, Fayetteville, Arkansas 72701, USA}
\author{Kyungwha Park}
\email{kyungwha@vt.edu}
\affiliation{Department of Physics, Virginia Tech, Blacksburg, Virginia 24061, USA}

\date{\today}

% our key findings:
% 1. Topological nodal structure for arbitrary B-field direction: (a) Nodal line for B field along a or b axis. (b) The rest: Weyl.
% (c) k.p model: needs higher-order terms for consideration of correct symmetry. (d) k.p model: general form of nodal structure for
% arbitrary B field direction.
% 2. Anomalous Hall conductivity sigma_{ac} as a function of B field direction.

\begin{abstract}
We investigate the topological nodal structure of three-dimensional (3D) ZrTe$_5$ driven by Zeeman splitting as a function of the direction
of external magnetic ({\bf B}) field by using a Wannier-function-based tight-binding (WFTB) model obtained from first-principles calculations.
It is known that small external stimuli can drive 3D ZrTe$_5$ into different topological phases including Dirac semimetal. In order to emphasize
the effect of Zeeman splitting, we consider 3D ZrTe$_5$ in a strong TI phase with a small band gap. With Zeeman splitting greater than the
band gap, the WFTB model suggests that a type-I nodal ring protected by (glide) mirror symmetry is formed when the {\bf B} field aligns with
the crystal $a$ or $b$ axes, and that a pair of type-I Weyl nodes are formed otherwise, when conduction and valence bands touch.
We show that a pair of Weyl nodes can disappear through formation of a nodal ring, rather than requiring two Weyl nodes with opposite
chirality to come together. Interestingly, a type-II nodal ring appears from crossings of the top two valence bands when the {\bf B} field
is applied along the $c$ axis. This nodal ring gaps out to form type-II Weyl nodes when the {\bf B} field rotates in the $bc$ plane.
Comparing the WFTB and linearized $k \cdot p$ model, we find
inadequacy of the latter at some ${\bf B}$ field directions. Further, using the WFTB model, we numerically compute the intrinsic
anomalous Hall conductivity ${\sigma_{ac}}$ induced by Berry curvature as a function of chemical potential and {\bf B} field direction. We
find that ${\sigma_{ac}}$ increases abruptly when the {\bf B} field is tilted from the $a$ axis within the $ab$ plane. Our WFTB model
also shows significant anomalous Hall conductivity induced by avoided level crossings even in the absence of Weyl nodes.
%Our findings can be applied to other Dirac semimetals with mirror symmetries.
\end{abstract}
%\pacs{TBD}

\maketitle

\section{Introduction}

% motivation: ZrTe5 is unique in (i) close to the critical point of topological phase transition; (ii) quantum limit in a relatively small B field
% a few tesla; (iii) intriguing electron and thermal transport measurements

Among various three-dimensional (3D) topological materials, ZrTe$_5$ is unique in that small volume changes \cite{HWeng_PRX2014,ZFan_SciRep2017},
small strains \cite{Mutch2019,YZhang_NatComm2017}, or even moderate temperature \cite{BXu_PRL2018} can induce a topological phase transition
from weak to strong topological insulator (TI) phase. This unique aspect may be attributed to the fact that a single ZrTe$_5$ layer is a two-dimensional (2D) TI \cite{HWeng_PRX2014}. Before the experimental discovery of weak TI, $\beta$-type bismuth iodide (Bi$_4$I$_4$) \cite{Noguchi2019}, ZrTe$_5$ was a first realistic candidate for a weak TI. In order to drive the material from a weak to strong TI by external
means, the band gap must be closed at a critical value, which manifests Dirac semimetal \cite{BJYang2014}. However, this Dirac semimetal phase
is not protected by crystal rotational symmetry like Na$_3$Bi and Cd$_3$As$_2$ \cite{ZWang_PRB2012,ZWang_PRB2013,BJYang2014}
or by nonsymmorphic group symmetry like $\beta$-cristobalite BiO$_2$ \cite{SMYoung_PRL2012}. The sensitivity of the topological phase to
small changes of the lattice constant or temperature, places this material as an ideal playground for exploring effects of external stimuli
on topological properties. Furthermore, such sensitivity resulted in experimental observations of weak TI \cite{YZhang_NatComm2017,HXiong_PRB2017,YYLv_PRB2018}, strong TI \cite{YZhang_NatComm2017,GManzoni_PRL2016}, and Dirac semimetal phases \cite{JLZhang_PRL2017,RYChen_PRL2015,JWang_PNAS2018} in 3D ZrTe$_5$.

% The anomalous peak T_p in the longitudinal resistivity is due to a transition of carrier type from hole to electron as temperature decreases.
% Ong's exp: T_p is very close to zero, implying that mostly hole carriers. Chemical potential below the Fermi level for ideal ZrTe5.

% 2019 Nature paper: F. Tang et al., ``Three-dimensional quantum Hall effect and metal-insulator transition in ZrTe5,''
% experimentally observed 3D quantum Hall effect with CDW formation along the b axis
% and various quantum phase transitions driven by magnetic field
% T_p=95 K, v_F (in 10^5 m/s): 9+/-3 (a axis), 0.3+/-0.1 (b axis, this is the most important), 1.9+/-0.7 (c axis)

Recent magnetotransport experiments on 3D ZrTe$_5$ showed interesting features including a large anomalous Hall effect as a function of
the orientation of external magnetic ({\bf B}) field (despite the absence of magnetic order) \cite{Liang2018,JGe2019} as well as 3D
quantum Hall effect and metal-insulator transition \cite{FTang_Nature2019}. In general the intrinsic anomalous Hall effect requires
contributions of nonzero Berry curvature from occupied bands \cite{Di_Xiao2010,Nagaosa2010}.
When the {\bf B} field is rotated out of plane, the anomalous Hall resistivity was observed to abruptly increase in an antisymmetric
fashion or reveal strong asymmetry as a function of the field orientation \cite{Liang2018}. On the other hand, for the in-plane {\bf B} field,
the anomalous Hall resistivity was observed to show clear antisymmetry as a function of the field orientation \cite{Liang2018}. The latter
feature cannot be explained by the planar Hall effect ~\cite{Nandy_PRL2017,Burkov2017} alone. Theoretical efforts have been so far mostly
limited to understanding topological nodal structures using lowest-order effective models when the {\bf B} field is parallel to the crystal
axes \cite{RYChen_PRL2015}. Very recently, Burkov \cite{Burkov2018} proposed an effect of mirror anomaly on the intrinsic anomalous Hall
conductivity (AHC) for Dirac semimetals when the {\bf B} field rotates. The anomalous Hall effect observed in Ref.~\cite{Liang2018} has
not been theoretically understood yet.

In order to provide insight into the origin of the intriguing anomalous Hall effect, we construct a Wannier-function-based tight-binding (WFTB)
model for 3D ZrTe$_5$ from first-principles calculations and investigate topological phase transitions induced by
Zeeman splitting while ignoring Landau levels. The magnitude of the {\bf B} field is fixed such that the Zeeman splitting is greater than a small
band gap, while the {\bf B} field direction is varied within the crystal $ab$, $bc$, and $ac$ planes. The WFTB model predicts that a pair of
type-I Weyl nodes are formed for any direction of {\bf B} field except for when the {\bf B} field is parallel to the $a$ or $b$ axes, considering crossings of the conduction and valence bands. This pair of Weyl nodes abruptly transforms into a nodal ring when the {\bf B} field
aligns with the $a$ or $b$ axis, which conceptually differs from annihilation of Weyl nodes with opposite chirality at the same
$k$ point. Interestingly, when the top two valence bands cross, the WFTB model suggests type-II topological nodal structures depending on the direction of {\bf B} field. We also show that the linearized $k \cdot p$ model is not enough to capture even qualitatively correct nodal
structures for some {\bf B} field directions. We numerically compute the intrinsic AHC as a function of the orientation of {\bf B} field and
chemical potential, using the WFTB model. Our results can be compared with the experimental antisymmetric component
of the intrinsic out-of-plane AHC as a function of the tilting angle.

% Outline of the paper for PRB:

We present the crystal structure and symmetries of ZrTe$_5$ in Sec.~\ref{sec2} and construction of the WFTB model in Sec.~\ref{sec3}. We show the calculated band structures using the WFTB model in the presence of {\bf B} field and discuss the induced topological phases as a function of
{\bf B}-field direction in Sec.~\ref{sec4}. We compare our findings from the WFTB model with those from the linearized $k \cdot p$ model in Sec.~\ref{sec5}. Then we present and analyze the calculated AHC as a function of {\bf B}-field direction and chemical potential in Sec.~\ref{sec6}.
We summarize our conclusions in Sec.~\ref{sec7}.

%From Jack: We confirm the topological charge of the Weyl nodes by calculating the Berry Curvature flux.

\section{Crystal Structure and Symmetries}\label{sec2}

\subsection{Crystal structure}

\begin{figure}[htb]
\centering
\includegraphics[width=0.45 \textwidth]{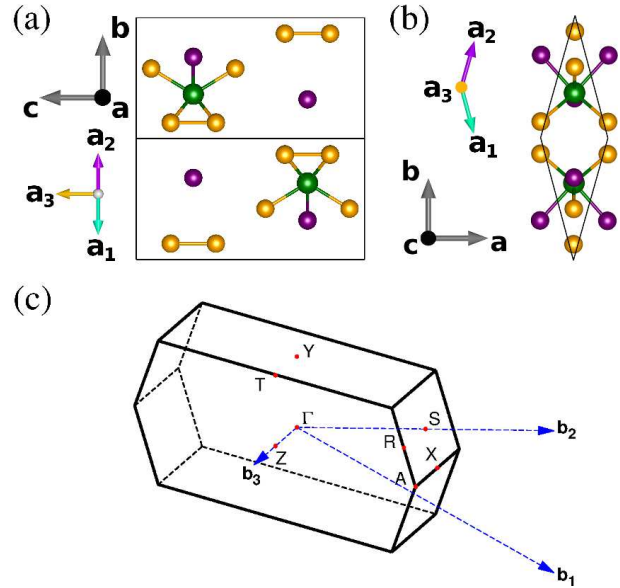}
\caption[geofig]{(a)-(b) Side views of ZrTe$_5$ unit cell. Zr atoms at Wyckoff position $4c$ are green, and Te atoms at $4c$ ($8f$) are purple (orange).
 Here ${\bf a}_1$, ${\bf a}_2$, and ${\bf a}_3$ are the primitive cell Bravais lattice vectors. (c) First BZ with high-symmetry $k$ points and reciprocal lattice vectors shown.}
\label{fig:geo}
\end{figure}

Bulk ZrTe$_5$ crystallizes in the orthorhombic structure with space group $Cmcm$ (No. 63), $D_{2h}$, where the experimental
lattice constants are $a=3.9797$, $b=14.470$, and $c=13.676$~\AA~\cite{Fjellvag1986}.
A primitive unit cell [Fig.~\ref{fig:geo}(a) and (b)] contains two Zr
and ten Te atoms. The Zr atoms (green) are located at Wyckoff position $4c$, and the two Te atoms (purple) and eight Te atoms (orange) are at $4c$ and $8f$, respectively.
Each 2D zigzag layer connected along the $a$ and $c$ axes is well separated by $\frac{b}{2}$ and stacked along the $b$ axis with weak van der Waals interaction.
Each Zr atom is bonded with eight Te atoms [Fig.~\ref{fig:geo}(b)]. We consider the following Bravais lattice vectors for the primitive
unit cell: {\bf a}$_1$=($\frac{a}{2}$,$-\frac{b}{2}$,0), {\bf a}$_2$=($\frac{a}{2}$,$\frac{b}{2}$,0), {\bf a}$_3$=(0,0,$c$)
in Cartesian coordinates.
In our convention, the $a$, $b$, and $c$ axes are the $x$, $y$, and $z$ axes in Cartesian coordinates.
The corresponding reciprocal lattice vectors are:
{\bf b}$_1$=2$\pi$($\frac{1}{a}$,$-\frac{1}{b}$,0), {\bf b}$_2$=2$\pi$($\frac{1}{a}$,$\frac{1}{b}$,0), and {\bf b}$_3$=2$\pi$(0,0,$\frac{1}{c}$).
The first Brillouin zone (BZ) is shown in Fig.~\ref{fig:geo}(c), where $S$=(0,$\frac{1}{2}$,0), $\Gamma$=(0,0,0), $Z$=(0,0,$\frac{1}{2}$), $R$=(0,$\frac{1}{2}$,$\frac{1}{2}$), $Y$=($-\frac{1}{2}$,$\frac{1}{2}$,0), and $T$=($-\frac{1}{2}$,$\frac{1}{2}$,$\frac{1}{2}$), in fractional coordinates.
These high-symmetry $k$ points are equivalent to $X$, $\Gamma$, $Y$, $M$, $Z$, and $R$ in Ref.\cite{HWeng_PRX2014}, respectively. The zone boundary point $X$ is located at ($\eta$,$\eta$,0), where $\eta=\frac{1}{4}(1+\frac{a^2}{b^2})$.

In order to examine the effect of Zeeman splitting, we apply 0.25\% compressive uniaxial stress along the $b$ axis to the DFT-relaxed unstrained geometry while keeping the volume fixed, such that ZrTe$_5$ remains in a strong TI phase with a small band gap in the presence of spin-orbit-coupling (SOC). For reference, the relaxed unstrained lattice constants are $a=4.0341$, $b=14.6998$, and $c=13.8843$~\AA.~In the strained case, the lattice constants are $a=4.0391$, $b=14.6630$, and $c=13.9017$~\AA.
The results obtained from the WFTB model correspond to the 0.25\% strained structure.

\subsection{Symmetries}\label{sec2:symmetries}

3D ZrTe$_5$ has inversion symmetry as well as the following crystal symmetries: two-fold rotational symmetries along the $a$ and $b$ axes ($C_{2a}$ and $C_{2b}$), two-fold screw symmetry along the $c$ axis, mirror symmetries about the $ab$ and $bc$ planes ($M_{ab}$, $M_{bc}$), and glide mirror symmetry about
the $ac$ plane ($M_{ac}$). Since the inversion center does not coincide with the origin of the rotational symmetries, the space
group is nonsymmorphic. Inversion symmetry persists even in the presence of {\bf B} field. Depending on the direction of {\bf B} field,
the following symmetries can survive: (screw) $C_2$ symmetry about the {\bf B} field direction and the (glide) mirror symmetry about the plane
perpendicular to the {\bf B} field, or $C_{2\perp}{\cal T}$ where $C_{2\perp}$ is $C_2$ symmetry about the direction perpendicular
to the {\bf B} field, and ${\cal T}$ is the time-reversal operator.

\section{Construction of Wannier-function tight-binding model}\label{sec3}

We first calculate the electronic structure of bulk strained ZrTe$_5$ without SOC and ${\bf B}$ field by using the density-functional theory (DFT) code
{\sc VASP} \cite{VASP1,VASP2}. With the DFT-calculated band structure and initial atomic orbitals, we generate Wannier functions (WFs) by using
{\sc Wannier90} \cite{Wannier90}. Then we construct a SOC-free tight-binding model from the WFs and add atomic-like SOC to the tight-binding
model such that the model-calculated band structure agrees with the DFT result. Last, we add Zeeman energy to the tight-binding model.

\subsection{Initial DFT calculations}\label{sec3:DFT}

We perform the DFT calculations using {\sc VASP} \cite{VASP1,VASP2} within the Perdew-Burke-Ernzerhof (PBE) generalized-gradient approximation
(GGA) \cite{Perdew1996} for the exchange-correlation functional with and without SOC.  We use projector augmented wave (PAW) pseudopotentials \cite{Bloch1994} with an energy cutoff of 350 eV and a $19 \times 19 \times 5$ Monkhorst-Pack $k$-point mesh. For the experimental geometry~\cite{Fjellvag1986}, our DFT calculation shows that bulk ZrTe$_5$
with SOC is in a strong TI phase with a direct band gap of about 100~meV.
The structure with 0.25\% compressive strain along the $b$ axis has a band gap of 2.2~meV.
All calculated band structures from the WFTB model correspond to the strained structure, unless specified otherwise.

\subsection{SOC-free Hamiltonian}\label{sec3:spinfree}

In order to construct the SOC-free WFTB model, we start with an initial set of 40 projected atomic orbitals comprised of
$d_{z^2}$, $d_{x^2-y^2}$, $d_{xy}$, $d_{yz}$, and $d_{xz}$ orbitals centered at two Zr sites and $p_{x}$, $p_{y}$, and $p_{z}$ orbitals
centered at ten Te sites in the primitive unit cell. We need to include both Zr $d$ orbitals and Te $p$ orbitals in the WFTB model
due to their large contributions to valence and conduction bands, respectively, near $\Gamma$ indicated by our DFT calculations.
With the VASP-calculated Bloch eigenvalues and eigenstates, we compute the overlap
matrix and projection matrix at each DFT-sampled $k$ point by using {\sc Wannier90} \cite{Wannier90,Marzari1997,Souza2001}.
We apply the disentanglement procedure within the outer energy window $[-8.23, 5.27]$~eV relative to the Fermi level.
In this energy window, the number of Bloch bands ranges from 43 to 47, and both occupied and unoccupied bands are included.
We check that the generated WFs are close to pure atomic orbitals with only real components. In order to maintain the features of the atomic orbitals, maximal localization is not applied in the wannierization.

Now we construct the SOC-free WFTB model by using the generated WFs, $|{\mathbf R}+{\mathbf s}_{\beta}\rangle$, centered at
${\mathbf R}+{\mathbf s}_{\beta}$, where ${\mathbf R}$ are Bravais lattice vectors and ${\mathbf s}_{\beta}$ denote the sites of
orbital ${\beta}$ ($\beta$$=$1,...,40). The SOC-free Hamiltonian matrix ${\cal H}_0$ \cite{Marzari1997} reads
\begin{eqnarray}
{\cal H}_{0,\alpha \beta}({\mathbf k}) &=& \langle \psi_{{\mathbf k},\alpha}| {\cal H}_0 |\psi_{{\mathbf k},\beta}\rangle,  \\
 &=& \sum_{{\mathbf R}} e^{-i {\mathbf k} \cdot ({\mathbf R}+{\mathbf s}_{\alpha}-{\mathbf s}_{\beta})}
 t_{\alpha \beta}({\mathbf R}-{\mathbf 0}), \label{eq:Hab} \\
t_{\alpha \beta}({\mathbf R}-{\mathbf 0}) &=& \langle {\mathbf R} + {\mathbf s}_{\alpha}| {\cal H}_0 | {\mathbf 0} + {\mathbf s}_{\beta} \rangle,
\end{eqnarray}
where $|\psi_{k}({\mathbf r})\rangle$ are Bloch states over the crystal momentum ${\mathbf k}$ space. Here $t_{\alpha \beta}({\mathbf R}-{\mathbf 0})$
is a hopping or tunneling parameter from orbital $\beta$ at site ${\mathbf s}_{\beta}$
in the home cell at ${\mathbf R}={\mathbf 0}$ to orbital $\alpha$ at site ${\mathbf s}_{\alpha}$ in the unit cell located at ${\mathbf R}$.

\subsection{Addition of SOC and Zeeman energy}

We add on-site SOC to the home-cell terms since the generated WFs are close to the atomic orbitals that we project onto. Considering the
spin degrees of freedom, the number of WFs (basis set functions) is now 80. The SOC Hamiltonian
becomes ${\cal H}_{\small SOC} = \lambda \mathbf{L}\cdot {\mathbf{\sigma}}$, where $\lambda$ is the SOC parameter, ${\mathbf L}$ is the orbital
angular momentum, and ${\mathbf {\sigma}}$ represent Pauli spin matrices. We find that
with $\lambda_{\small \rm{Zr}}=-0.12$~eV and $\lambda_{\small \rm{Te}}=0.60$~eV, the WFTB calculated band structure agrees well with the VASP-calculated band structure, especially near the $\Gamma$ point in the vicinity of the Fermi level.

% YLiu2016: Y. Liu et al., ``Zeeman splitting and dynamical mass generation in Dirac semimetal ZrTe5,'' Nature Comm. 12516 (2016).
% v_F (in 10^5 m/s): 1.7 (b-c plane), 5.2 (a-c plane), 2.2 (a-b plane).

We add the {\bf B} field as a Zeeman term only and do not include the Peierls phase in the hopping parameters $t_{\alpha \beta}$ because Landau
levels are ignored in our calculations. Then the Zeeman interaction reads
${\cal H}_{Z} = g \mu_{\mathrm B} \mathbf{S} \cdot \mathbf{B}$, where $g$ is the effective electronic $g$ factor,
$\mu_B$ is Bohr magneton and ${\mathbf S}$ is the spin angular momentum. The WFTB model has the following Hamiltonian:
\begin{eqnarray}
{\cal H} &=& {\cal H}_0 + {\cal H}_{\small SOC} + {\cal H}_Z.
\label{eq:Htot}
\end{eqnarray}
Hereafter we consider the Zeeman interaction energy of 10~meV unless specified otherwise. Experimental data
on ZrTe$_5$ indicates that the $g$ factor is highly anisotropic. The $g$ factor when {\bf B} field is along the $b$ axis ($g_y$) is 21.3 \cite{YLiu2016,RYChen_PRL2015}, whereas the $g$ factor for the $a$ axis ($g_x$) is about 3.19 \cite{YLiu2016}.
The $g$ factor for the $c$ axis ($g_z$) was not reported. Therefore, for {\bf B} field along the $b$ axis, the Zeeman energy
of 10~meV corresponds to a {\bf B} field of about 16~T, considering that the Zeeman energy can be expressed as
$\sqrt{g_x^2 B_x^2 + g_y^2 B_y^2 + g_z^2 B_z^2} \frac{\mu_B}{2}$. As long as the Zeeman energy is greater than the band gap, the topological
phase transitions presented in this work can be realized. For the strained case, the band gap of 2.2~meV implies a requisite minimum {\bf B} field of 3.6~T when the {\bf B} field is applied along the $b$ axis.

\subsection{Comparison of WFTB-calculated to DFT-calculated band structure with and without SOC}\label{sec3:WFTBSOC}

% unstrained DFT band structure in Appendix?

\begin{figure}[htb]
\begin{center}
\includegraphics[width=0.4 \textwidth]{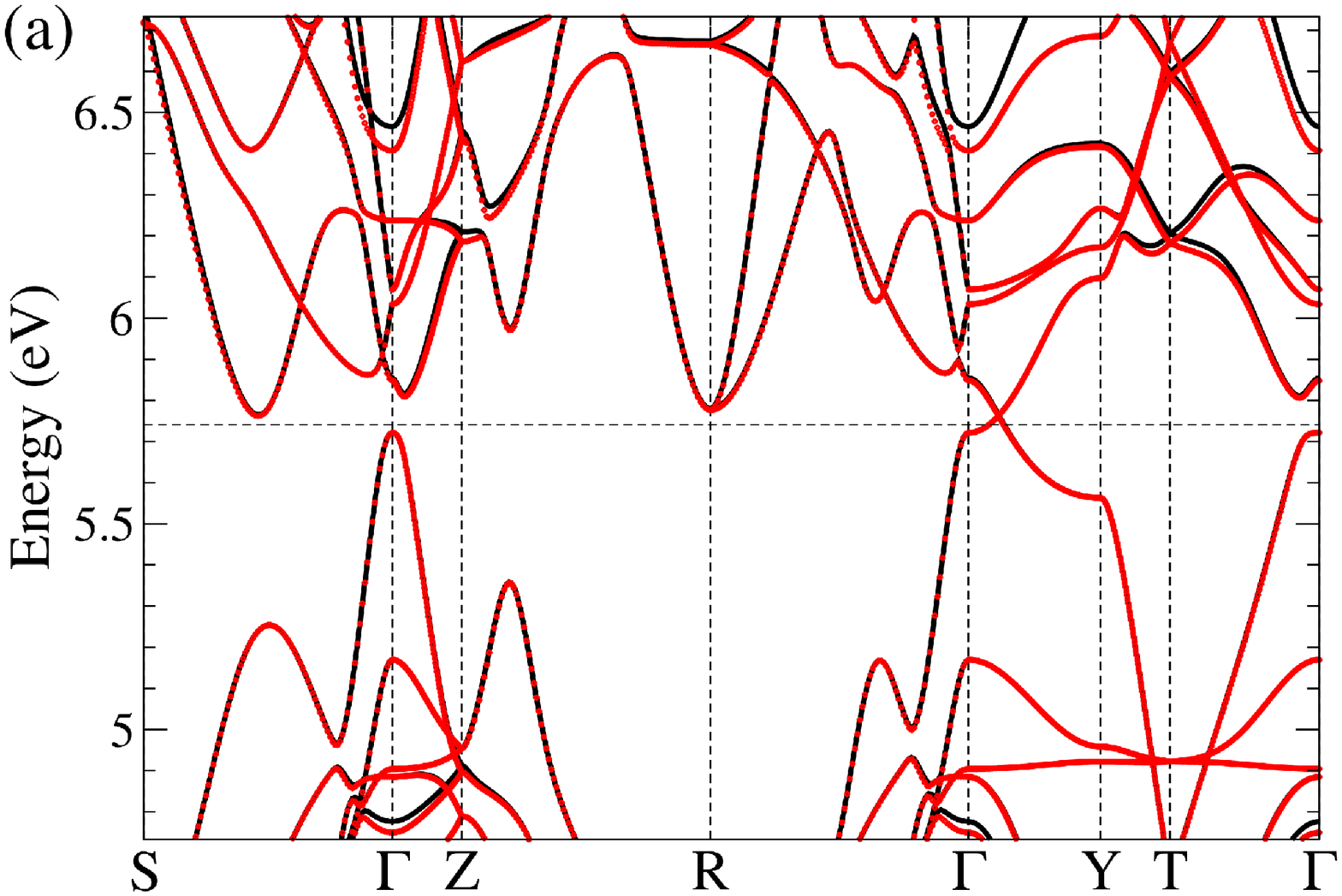}
\hspace{0.5truecm}
\includegraphics[width=0.4 \textwidth]{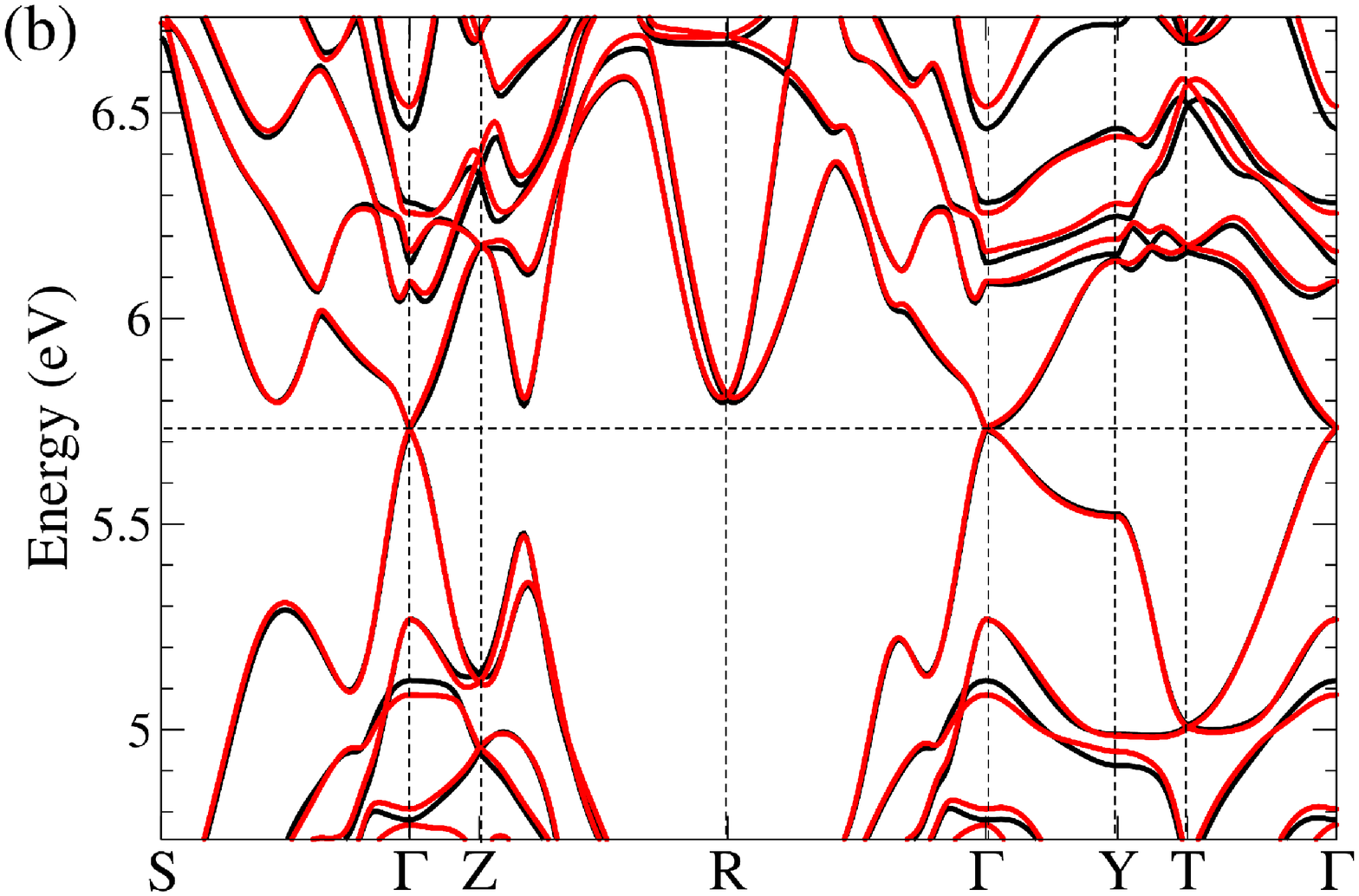}
\caption[a figure]{(a) Comparison between the VASP-calculated (black) and the WFTB-calculated (red) band structures for strained ZrTe$_5$
without SOC in the primitive unit cell. (b) Likewise with SOC included. The dashed lines indicate the Fermi levels. No {\bf B} field is included here.}
\label{fig:bulk}
\end{center}
\end{figure}

%We diagonalize the $80 \times 80$ matrix, ${\cal H}_0+{\cal H}_{\small SOC}$, with the same $k$ points and compare the eigenvalues XXXX.

In the absence of the Zeeman term, we compare the DFT-calculated band structure of bulk strained ZrTe$_5$ to the WFTB-calculated result
with and without SOC, as shown in Fig.~\ref{fig:bulk}. Without SOC, the valence and conduction bands meet linearly at a Dirac point along
the $\overline{\Gamma Y}$ direction or $\pm y$ axis. The WFTB-calculated band structure agrees well with the DFT result up to about
$\pm$1.0 eV from the Fermi level. With SOC, a small band gap of 2.2 meV opens up at $\Gamma$ which is also reliably captured by the WFTB model.
We calculate the 3D topological indices ($\nu_0$; $\nu_1$,$\nu_2$,$\nu_3$) \cite{LiangFu2007} of strained ZrTe$_5$ using the DFT-calculated wave function. For the reciprocal
vector ${\mathbf G}=\nu_1{\mathbf b}_1 + \nu_2{\mathbf b}_2 + \nu_3{\mathbf b}_3$, we find that ($\nu_0$; $\nu_1$,$\nu_2$,$\nu_3$)=(1;110).
Since $\nu_0=1$, strained ZrTe$_5$ is a strong TI. Each band is at least doubly degenerate due to the inversion and time-reversal
symmetries. In the $k_z=\frac{\pi}{2}$ plane, the four time-reversal invariant $k$ points are fourfold degenerate due to the additional mirror
symmetry $M_{ab}$.
%We check that the WFTB model calculated wave function also has the same 3D topological indices and the mirror eigenvalues as well
%as the same rotational symmetries as the DFT-calculated wave function.
%As shown in Appendix, we find that ZrTe$_5$ with the unstrained experimental geometry is also a strong TI.

\section{Zeeman-splitting induced topological phases from WFTB model}\label{sec4}

%\subsection{$k \cdot p$ model} in appendix??

In the presence of {\bf B} field, we diagonalize the $80 \times 80$ Hamiltonian matrix, Eq.~(\ref{eq:Htot}), with the same $k$ points as the
DFT calculation. We examine the topological properties and evolution of the nodal structure as a function of {\bf B}-field direction for a fixed
Zeeman energy of 10~meV with an isotropic $g$ factor ($g=2.0$) for simplicity. We consider cases that the {\bf B} field is applied along
the $a$, $b$, and $c$ axes as well as in the $ab$, $bc$, and $ac$ planes. The energy window of interest is [-0.12,+0.05]~eV relative to the
Fermi level $E_{\rm F}$, considering that the bulk ZrTe$_5$ samples studied in Ref.~\cite{Liang2018} are slightly hole-doped. Within this
energy window, the following three types of gapless crossings are in principle possible:
crossings between the bottom two conduction bands, crossings between the bottom conduction and top valence bands, and crossings
between the top two valence bands. However, we do not find crossings between the bottom two conduction bands for any {\bf B}-field directions.
We focus on the latter two types of crossing only.

\subsection{Magnetic field along the $a$ axis: Nodal-ring semimetal}\label{sec4:aaxis}

\begin{figure}[htb]
\begin{center}
%\vspace{0.5truecm}
\includegraphics[width=0.2 \textwidth]{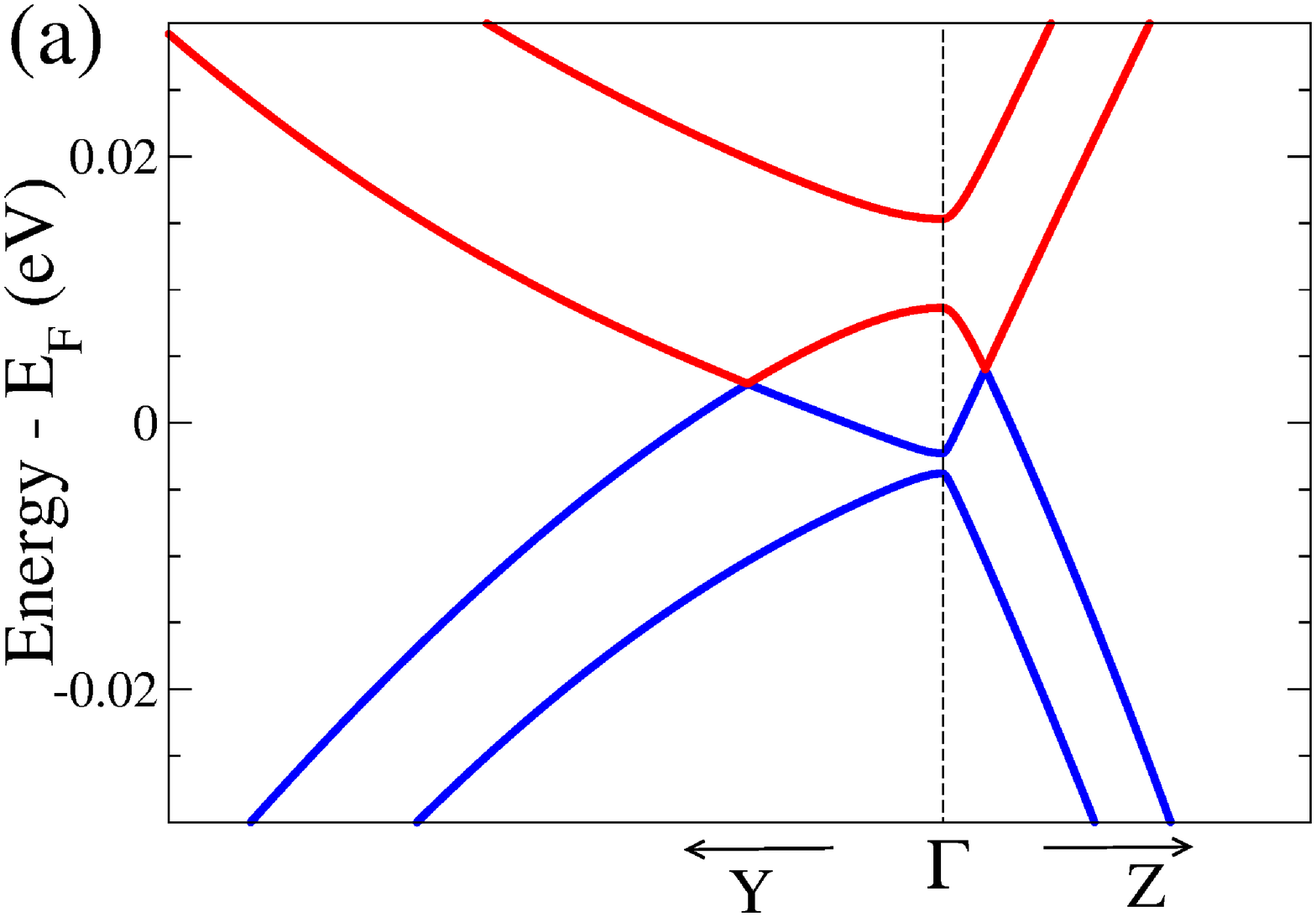}
\hspace{0.2truecm}
\includegraphics[width=0.25 \textwidth]{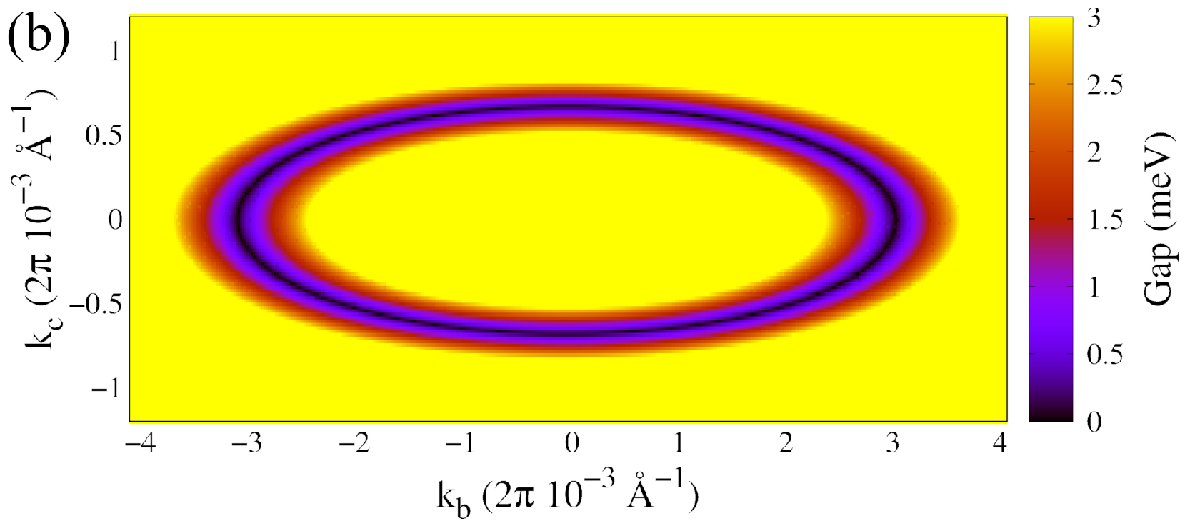}
\caption[a figure]{(a) WFTB-calculated band structure relative to the Fermi level $E_{\rm F}$ along the $Y-{\Gamma}-Z$ direction when the
{\bf B} field aligns with the $a$ axis. (b) The corresponding gapless nodal ring in the $k_b$-$k_c$ plane with the energy gap in color scale.}
\label{fig:Ba-nodal}
\end{center}
\end{figure}

Figure~\ref{fig:Ba-nodal}(a) shows the WFTB-calculated band structure along the $Y-\Gamma-Z$ direction near $\Gamma$ in the vicinity of the Fermi
level, when the {\bf B} field is applied along the $a$ axis. The gapless points are found from crossings between the top valence
and bottom conduction bands in the $k_b$-$k_c$ plane and they form a ring, as shown in Fig.~\ref{fig:Ba-nodal}(b). Since the two bands meet
with opposite slope, this is a type-I nodal ring. Note that with the {\bf B}
field, the relevant remaining symmetries are inversion symmetry, $C_{2a}$ and $M_{bc}$. From the eigenvectors of the WFTB
model, we confirm that the two crossing bands have opposite $M_{bc}$ mirror eigenvalues. The intercepts with the $b$ and $c$ axes are
(0, $\pm$0.003072, 0) and (0, 0, $\pm$0.000673)~$2\pi\cdot$\AA$^{-1}$, respectively.
The gapless ring is not an equi-energy curve, and no other gapless points are found within the energy window of interest.

In order to identify the topological nature of the gapless ring, we compute the Berry phase ${\varphi}_{\rm B}$ around a closed circle ${\cal C}$
interlocking the gapless ring. The Berry phase is defined as a sum of line integrals of the Berry connection of all occupied bands $n$,
${\mathbf A}_n({\mathbf k})$, over a closed path ${\cal C}$ in $k$ space:
\begin{equation}
\varphi_{\rm B} = \sum_{n=1}^{\rm{occ}} \oint_{{\cal C}} d{\bf k} \cdot {\mathbf A}_n({\mathbf k})
\end{equation}
where ${\mathbf A}_n({\mathbf k})=i \langle u_{n{\mathbf k}}|{\mathbf \nabla_{\mathbf k}}u_{n{\mathbf k}}\rangle$. Here $u_{n{\mathbf k}}$
is a periodic function of the Bloch state. We find that the Berry phase is $\pi$. Thus, the ring of the gapless points is indeed
a topological nodal ring.

\subsection{Magnetic field along the $b$ axis: Nodal-ring semimetal}\label{sec4:baxis}

% KP: dominant contributions from Zr dyz and d_{x^2-y^2} for the valence and conduction bands strained with SOC.

\begin{figure}[htb]
\begin{center}
%\vspace{0.6truecm}
\includegraphics[width=0.24 \textwidth]{Fig4a_Sep02_2019.eps}
\hspace{0.2truecm}
\includegraphics[width=0.21 \textwidth]{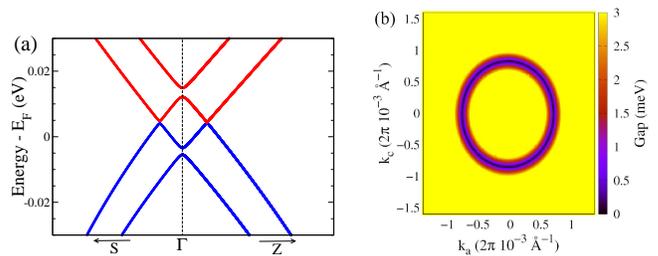}
\caption[a figure]{(a) WFTB-calculated band structure along the $S-\Gamma-Z$
directions when {\bf B} field is parallel to the $b$ axis. (b) The corresponding gapless nodal ring in the $k_a$-$k_c$ plane.}
\label{fig:Bb-nodal}
\end{center}
\end{figure}

Figure~\ref{fig:Bb-nodal}(a) shows the calculated band structure along the $S-\Gamma-Z$ direction when the {\bf B} field aligns with the $b$ axis.
The bottom conduction and top valence bands meet near the Fermi level along the $\Gamma-Z$ and $\Gamma-X$ directions (not shown),
whereas a small gap opens up along the $\Gamma-S$ direction. Similar to Sec.~\ref{sec4:aaxis}, a type-I ring of gapless points is found in the $k_a$-$k_c$
plane near $\Gamma$ in the vicinity of the Fermi level [Fig.~\ref{fig:Bb-nodal}(b)]. The gapless points intercept the
$k_a$ and $k_c$ axes at ($\pm$0.000743, 0, 0) and (0, 0, $\pm$0.000840)~$2\pi\cdot$\AA$^{-1}$.
We expect that the gapless crossings are allowed because the two crossing bands have opposite $M_{ac}$ glide mirror eigenvalues.
We find that the Berry phase is $\pi$; the gapless ring
is a topological nodal ring. No other gapless crossings are found in the energy window of interest.

%% KP August 7 2019: Need to decide if we need this paragraph here or discuss this only in AHC section.
%There are two distinct features in this {\bf B} field compared to Case I. Firstly, we find that the top valence and bottom conduction
%bands open up a small gap of 0.5~meV at the same energy (almost the Fermi level) along the $S-\Gamma$ and $R-\Gamma$ directions %[Fig.~\ref{fig:Bb-nodal}(a)]
%{\bf (KP: This needs to be checked with Jack's correct 3newparallel.f code)}. These avoided level crossings with a small gap are expected to greatly %contribute to the anomalous Hall conductivity, despite the absence of the topological charge associated with the crossings.

\subsection{Magnetic field along the $c$ axis: Weyl or type-II nodal-ring semimetal}\label{sec4:caxis}

% We obtained v_F for the B field along the c axis or z axis near the Weyl point.
% Tilting of Dirac cone at the Weyl point along the c axis is very small like 11% difference in the Fermi velocity
% WFTB v_F= 2.55 vs 2.26 x 10^5 m/s along the k_z axis !! agrees well with v_F estimated from 2019 Nature 3D QHE paper (YLiu2019)
% WFTB v_F= 2.87 x 10^5 m/s along the k_x axis         !! agrees somewhat (a factor of 2 smaller) with v_F estimated from 2019 Nature 3D QHE paper (YLiu2019)
% WFTB v_F= 0.46 x 10^5 m/s along the k_y axis         !! agrees well with v_F estimated from 2019 Nature 3D QHE paper (YLiu2019)

\begin{figure}[htb]
\begin{center}
\includegraphics[width=0.47\textwidth]{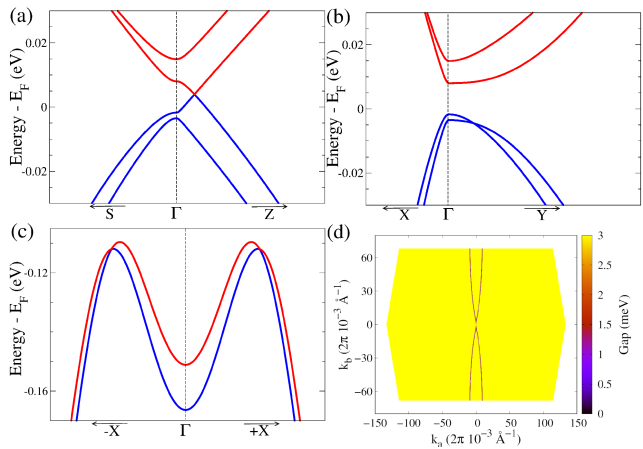}
\caption[Nodal lines]{(a)-(c) Band structures when {\bf B} field is parallel to the $c$ axis. For (c), $k_y$ is fixed to be
0.034100$~$2${\pi}$$\cdot$\AA$^{-1}$. (d) Type-II nodal ring in the $k_a$-$k_b$ plane with the gap size in color scale. Here
the yellow region indicates the whole first BZ.}
\label{fig:Bc-weyl}
\end{center}
\end{figure}

When the {\bf B} field aligns with the $c$ axis, the valence and conduction bands meet with opposite slope along the $c$-axis at
(0, 0, $\pm$0.000624)~2$\pi$$\cdot$\AA$^{-1}$  in the vicinity of the Fermi level as shown in Fig.~\ref{fig:Bc-weyl}(a).
We evaluate the topological charge $\chi_n$ associated with the gapless points by computing the
Berry curvature ${\mathbf \Omega}_n({\mathbf k})$ using the method discussed in {\sc WannierTools} \cite{WTOOLS} and Ref.~\cite{Villanova2018}.
The Berry curvature can be calculated as \cite{XWang2007}
\begin{widetext}
\begin{equation}
\epsilon_{\alpha\beta\gamma}\Omega_{n,\gamma}({\bf k}) =
-2 {\rm Im} \sum_{m \neq n} \frac{\langle\langle \phi_n({\bf k}) \|{\cal H}_{\alpha}\|\phi_m({\bf k}) \rangle\rangle
\langle\langle \phi_m({\bf k}) \|{\cal H}_{\beta}\|\phi_n({\bf k}) \rangle\rangle}{({\cal E}_m({\bf k}) - {\cal E}_n({\bf k}))^2},
\label{eq:omega}
\end{equation}
\end{widetext}
where ${\cal H}_{\alpha}\equiv \partial {\cal H} \slash \partial k_{\alpha}$. Here $\|\phi_n({\bf k}) \rangle\rangle$ and
${\cal E}_n({\bf k})$ are the $n$-th eigenvector and eigenvalue of ${\cal H}({\bf k})$ [Eq.~(\ref{eq:Htot})], and
$\epsilon_{\alpha\beta\gamma}$ is the Levi-Civita tensor without a sum over $\gamma$. The topological charge $\chi_n$ of
each gapless point (arising from a crossing of band $n$ and band $n+1$) is then calculated by enclosing it in
a small sphere ${\cal S}$,
\begin{equation}
\chi_n = \frac{1}{2\pi}\oint_{{\cal S}} \sum_{l=1}^{n} dS \ \hat{\mathbf{n}} \cdot \mathbf{\Omega}_l(\mathbf{k}),
\label{eq:chi}
\end{equation}
where ${\mathbf{n}}$ is a unit vector normal to ${\cal S}$. We find that the topological charge associated with the gapless points are
$\chi_n=\mp 1$, respectively, and so they are type-I Weyl points. The two Weyl points are related by inversion symmetry.

In addition, we find that the top two valence bands touch in the vicinity of the Fermi level, as shown in Fig.~\ref{fig:Bc-weyl}(b) and (c).
The gapless points form a nodal ring in the $k_a$-$k_b$ plane [Fig.~\ref{fig:Bc-weyl}(d)]. There are two interesting features of this ring.
First, the band dispersion near the gapless point has the same slope along the $k_a$ and $k_b$ axes, and so it is referred to as a type-II nodal ring. Note that the nodal ring discussed earlier in Secs.\ref{sec4:aaxis} and \ref{sec4:baxis} is of type-I. Second, the type-II nodal ring
extends to the neighboring BZ, forming a closed thin cigar shape. The size of the type-II nodal ring is much larger than the type-I
nodal rings discussed in Secs.\ref{sec4:aaxis} and \ref{sec4:baxis}. We evaluate the Berry phase around the circle interlocking the ring,
finding that it is indeed $\pi$; the ring is topologically protected by $M_{ab}$ symmetry.

\subsection{Magnetic field in the $ab$ plane}\label{sec4:abplane}

\begin{figure*}[htb]
\begin{center}
\includegraphics[width=0.7\textwidth]{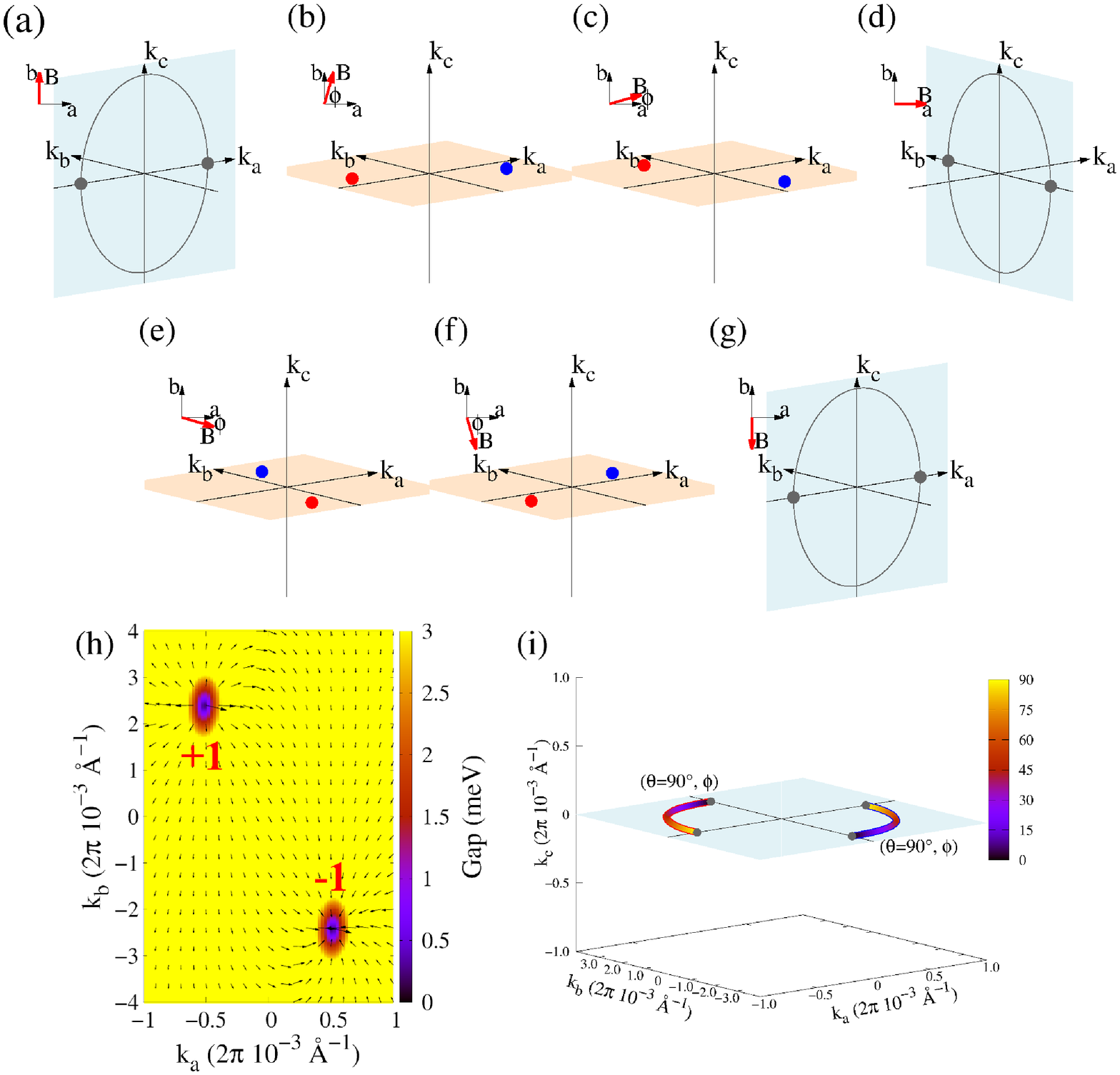}
\caption[Bab]{(a)-(g) Schematic evolution of the nodal ring and Weyl points as a function of the angle between {\bf B} field and the
$a$ axis, $\phi$, (h) WFTB-calculated Berry curvature in the $k_a$-$k_b$ plane at $\phi=45^{\circ}$, and (i) WFTB-calculated evolution
of the nodal structure as a function of $\phi$, when the {\bf B} field is in the $ab$ plane.
In (a), (d), and (g), the blue plane indicates the mirror-symmetry plane where the topological nodal ring resides.
In (b), (c), (e), and (f), the red and blue filled circles correspond to Weyl points with topological charge $\chi$ of $+1$ and $-1$,
respectively. In (h), the topological charge of the Weyl points is denoted, and the energy gap is shown in color scale.
In (i), the color scale indicates the value of $\phi$.}
\label{fig:Bab}
\end{center}
\end{figure*}

Figure~\ref{fig:Bab}(a)-(g) schematically shows the evolution of the type-I nodal ring and Weyl points as the {\bf B} field rotates in the $ab$
plane. When the {\bf B} field is slightly rotated from the $b$ axis in the $ab$ plane, the WFTB model shows that the type-I nodal ring in the $k_a$-$k_c$ plane is abruptly gapped out (due to broken mirror symmetry) everywhere but two gapless points, transforming into a pair of
type-I Weyl nodes ($\chi_n={\pm}1$) lying in the $k_a$-$k_b$ plane. As the polar angle $\phi$ between the {\bf B} field and the $a$ axis further decreases, the Weyl points initially close to the $k_a$ axis evolve toward the $k_b$ axis in the $k_a$-$k_b$ plane. Then when the {\bf B}
field aligns with the $a$ axis ($\phi=0$), the pair of Weyl points suddenly transforms into a nodal ring in the $k_b$-$k_c$ plane (with the restoration of a mirror symmetry).
As the {\bf B} field continues to rotate clockwise beyond the positive $a$ axis ($\phi < 0$), the nodal ring transforms into a pair of
Weyl points where the chirality of the Weyl points is now exchanged.
For example, Fig.~\ref{fig:Bab}(h) exhibits the calculated Berry curvature with a pair of Weyl nodes obtained from the WFTB model at
${\phi}=45^{\circ}$. Figure~\ref{fig:Bab}(i) summarizes the calculated evolution of the nodal ring and Weyl points as a function of $\phi$
for 0 $\le$ $\phi$ $\le$ $90^{\circ}$, where the intercepts with the $k_a$ and $k_b$ axes become part of the nodal rings (not drawn) when
$\phi=90^{\circ}$ and $\phi=0$, respectively.

In contrast to a common belief, our result demonstrates that there is another way to annihilate Weyl points other than bringing a pair
of Weyl points with opposite chirality to the same $k$ point. This possibility was earlier discussed
within effective models in the case of Dirac semimetals in the presence of {\bf B} field~\cite{Burkov2018}. Furthermore, our finding
indicates that the chirality of the Weyl nodes can be exchanged with the reversal of the $b$ component of the {\bf B} field.

\subsection{Magnetic field in the $ac$ or $bc$ plane}\label{sec4:bcplane}

\begin{figure}[htb]
\begin{center}
\includegraphics[width=0.47\textwidth]{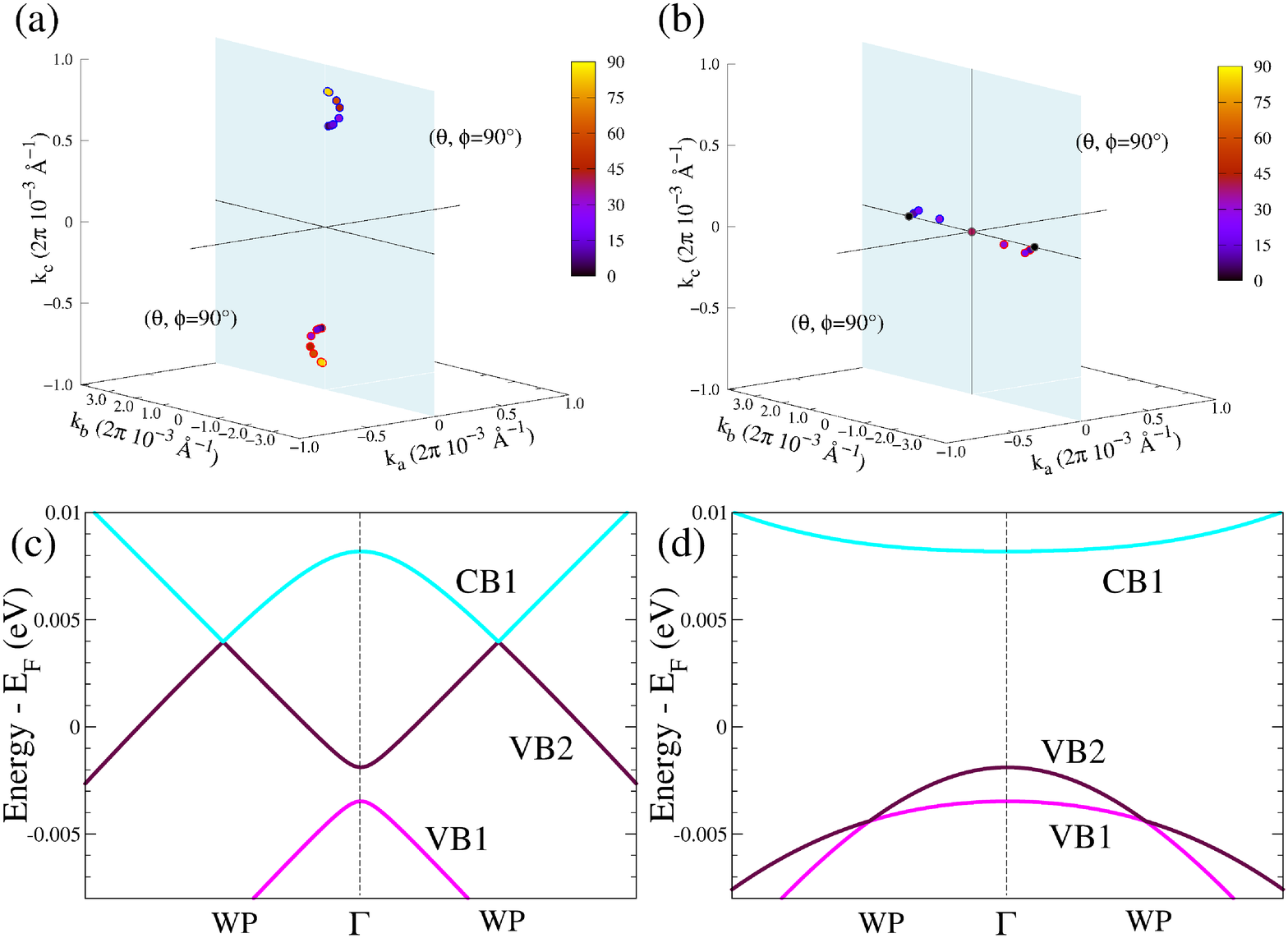}
\caption[Nodal lines]{(a)-(b) WFTB-calculated evolution of the type-I and type-II Weyl nodes, respectively, as a function of
$\theta$ (color scale) when the {\bf B} field rotates in the $bc$ plane. The type-I Weyl nodes arise from the
crossings of the conduction and valence bands, whereas the type-II Weyl nodes arise from the two valence bands.
Here $\phi$ is the angle between {\bf B} field and the $a$ axis. In (a) the Weyl points are located at
(0, $\pm$0.000540, $\mp$0.000757)~$2\pi{\cdot}$\AA$^{-1}$~for $\theta=45^{\circ}$.
(c)-(d) WFTB-calculated band structures near the crossings of the conduction (CB1) and valence bands (VB2) and
of the two valence bands (VB1 and VB2) for the {\bf B} field in the $bc$ plane with $\theta$=10$^{\circ}$.}
\label{fig:Bbc}
\end{center}
\end{figure}

When the {\bf B} field is rotated from the $c$ axis in the $bc$ plane, we find that the pair of type-I Weyl nodes with ${\chi}_n=\pm 1$
(arising from the conduction and valence bands) move somewhat away from the $k_c$ axis and return to the axis as the angle $\theta$
approaches $90^{\circ}$, where $\theta$ is the angle between the {\bf B} field and the $c$ axis. See Fig.~\ref{fig:Bbc}(a). Then when the
{\bf B} field aligns with the $b$ axis, the type-I Weyl points [Fig.~\ref{fig:Bbc}(c)] abruptly transform into the type-I nodal ring in the
$k_a$-$k_c$ plane, as discussed earlier. For $\theta > 90^{\circ}$, a pair of Weyl nodes reappear with the reversed topological charges
compared to those in the case of ${\theta} < 90^{\circ}$, similarly to Sec.~\ref{sec4:abplane}. The transformation of the Weyl nodes into a nodal
ring is related to the mirror anomaly~\cite{Burkov2018}, which is discussed later in Sec.~\ref{sec5:burkov}.

On the other hand, the type-II nodal ring arising from the top two valence bands is now completely gapped out except for two type-II gapless
points [Fig.~\ref{fig:Bbc}(d)] when the {\bf B} field is rotated from the $c$ axis in the $bc$ plane. For example, the type-II Weyl nodes occur
at (0, $\pm$0.002140, $\pm$0.000025)~$2\pi{\cdot}$\AA$^{-1}$~at $\theta=10^{\circ}$. We confirm that the type-II gapless
points have topological charge $\chi_n=\mp 1$, using the method discussed in Ref.~\cite{Soluyanov2015}. The type-II Weyl nodes with
opposite chirality are brought closer to each other as $\theta$ increases. They are eventually annihilated when $\theta$ approaches about 40$^{\circ}$ [Fig.~\ref{fig:Bbc}(b)].

When the {\bf B} field is rotated from the $c$ axis in the $ac$ plane, we find that the pair of type-I Weyl nodes with $\chi_n=\pm 1$ (arising from
the conduction and valence bands) remain almost along the $k_c$ axis with only slight changes in their locations. Then when the {\bf B} field is parallel to the $a$ axis, the Weyl nodes transform into a nodal ring. In contrast to the case with the {\bf B} field in the $bc$ plane,
there are no crossings from the top two valence bands in this case.

\subsection{Summary of topological phases from WFTB model}\label{sec4:summary}

Table~\ref{tab:summary} summarizes the topological phases found from the WFTB model. In the next section, we discuss the linearized
$k \cdot p$ model and the topological phases predicted from the $k \cdot p$ model. We also compare the findings
from the WFTB model and those from the $k \cdot p$ model.

% new table: KP, Sep16-2019
\begin{table*}[htb]
\centering
\caption{Zeeman-splitting driven topological phases in 3D ZrTe$_5$ as a function of the {\bf B}-field orientation based on the WFTB model and the linearized $k \cdot p$ model \cite{RYChen_PRL2015} (discussed in Sec.~\ref{sec5}), considering the energy window of [$-$0.12, 0.05] eV relative to the Fermi level. The second column corresponds to the nodal structure from the crossings of the bottom conduction and the top valence bands, and the third column for that of the top two valence bands in the WFTB model. In the case of the $k \cdot p$ model, the top two valence bands never meet. The topological phase for the {\bf B} field within the $ab$, $bc$, or $ac$ plane (marked by $*$) excludes the cases that the {\bf B} field coincides with the $a$, $b$, or $c$ axis. In the case of WFTB (VB-VB), the type-II Weyl nodes (marked by $^{\dag}$) are formed for only some angles ($< 40^{\circ}$) before they meet and annihilate at the same $k$ point, as in Fig.~\ref{fig:Bbc}(b).}
\begin{tabular}{|c|c|c|c|c|c|c|}
\hline \hline
Direction {\textbackslash} Model   & WFTB (CB-VB)        & WFTB (VB-VB)                  & $k \cdot p$ (CB-VB) \\ \hline
{\bf B}$\parallel${\bf a} & 1 type-I nodal ring &  -                            & 1 type-I nodal ring \\
{\bf B}$\parallel${\bf b} & 1 type-I nodal ring &  -                            & 1 type-I nodal ring \\
{\bf B}$\parallel${\bf c} & 2 type-I Weyl nodes & 1 type-II nodal ring          & 2 type-I Weyl nodes \\
{\bf B} in $ab$ plane*    & 2 type-I Weyl nodes &  -                            & 1 type-I nodal ring \\
{\bf B} in $bc$ plane*    & 2 type-I Weyl nodes & 2 type-II Weyl nodes$^{\dag}$ & 2 type-I Weyl nodes \\
{\bf B} in $ac$ plane*    & 2 type-I Weyl nodes &  -                            & 2 type-I Weyl nodes \\ \hline \hline
\end{tabular}
\label{tab:summary}
\end{table*}

\section{Comparison with linearized $k \cdot p$ model}\label{sec5}

\subsection{Lowest-order $k \cdot p$ model}\label{sec5:kdotp}

The lowest-order $k \cdot p$ Hamiltonian ${\cal H}_{\mathrm{kp}}({\bf k},{\bf B})$~\cite{RYChen_PRL2015} can be obtained by keeping only linear
terms in ${\bf k}$ that satisfy the symmetries of bulk ZrTe$_5$. The Hamiltonian ${\cal H}_{\mathrm{kp}}({\bf k},{\bf B})$ expanded near the $\Gamma$ point reads
\begin{eqnarray}
{\cal H}_{\mathrm{kp}}({\bf k},{\bf B}) &=& {\cal H}_{\mathrm{kp},0}({\bf k}) + {\cal H}_{\mathrm{kp},{\mathrm Z}}  \label{eq:Hkp} \\
{\cal H}_{\mathrm{kp},0}({\bf k}) &=& m \tau^z + v_x k_x \tau^x  \sigma^y + v_y k_y \tau^x \sigma^x +  v_z k_z \tau^y \\
{\cal H}_{\mathrm{kp},{\mathrm Z}} &=& \frac{1}{2} g \mu_B {\bf{\sigma}} \cdot {\bf B},
\end{eqnarray}
where $\tau^{x,y,z}$ and $\sigma^{x,y,z}$ are orbital (conduction and valence bands) and spin Pauli matrices. Here $v_{x,y,z}$ and $m$ are Fermi velocities and mass (or half of the bulk band gap), respectively. The $x$, $y$, and $z$ coordinates correspond to the crystal $a$, $b$, and $c$ axes.
${\cal H}_{\mathrm{kp},{\mathrm Z}}$ is the Zeeman interaction where we assume the same isotropic
$g$-factors for the conduction and valence bands for simplicity. The symmetry group of ZrTe$_5$ is generated by two mirror reflections,
$M_{ab}$, $M_{bc}$, inversion and time-reversal symmetries, which are represented by $- \tau^{z} \cdot i \sigma^{z}$,
$i \sigma^{x}$, $\tau^{z}$, and ${\cal K} \cdot i\sigma^{y}$, respectively (${\cal K}$ is complex conjugation).

Equation~(\ref{eq:Hkp}) can be diagonalized for an arbitrary {\bf B} field with energy eigenvalues given by
\begin{widetext}
\begin{equation}
\epsilon_{rs} ({\bf k},{\bf B}) = r \sqrt{m^2 + \left( \frac{g \mu_{B}}{2} \right)^2 B_0^2 + K^2
+ s \sqrt{m^2 B_0^2 + (v_z k_z)^2 B_0^2 + A_{\perp}^2 } g \mu_{B}},
\label{eq:Ekp}
\end{equation}
\end{widetext}
where $r,s=\pm$, $B_0^2= B_x^2 + B_y^2 + B_z^2$, $K^2 = (v_x k_x)^2+(v_y k_y)^2+(v_z k_z)^2 $, and $A_{\perp}^2 = (v_x k_x B_y + v_y k_y B_x)^2$.
In Ref.~\cite{RYChen_PRL2015} only three {\bf B} field directions ($a$, $b$, and $c$ axes) are considered.
The energy eigenvalues, Eq.~(\ref{eq:Ekp}), agree with those in Ref.~\cite{RYChen_PRL2015}.
We find that the conduction and valence bands can cross if the Zeeman splitting energy is greater than the band gap, i.e.,
$(\frac{g \mu_B B_0}{2})^2 > m^2$, whereas the two conduction (valence) bands can not cross each other for any {\bf B} field directions.
The resulting nodal structure is summarized in Table~\ref{tab:summary}.

\subsection{Comparison between WFTB and lowest-order $k \cdot p$ models}\label{sec5:kdotpcompare}

The nodal structure calculated using the WFTB and lowest-order $k \cdot p$ models qualitatively agree with each other in most cases, as
listed in Table~\ref{tab:summary}, although the positions of Weyl points or nodal rings may quantitatively differ from each other.
The previous study \cite{RYChen_PRL2015} reported the nodal structure only when the {\bf B} field aligns with the crystal axes using the lowest-order $k \cdot p$ model. {\it Qualitative} discrepancy between the WFTB model and the lowest-order $k \cdot p$ model occurs in two cases: (i) when the top two valence bands meet with each other for the {\bf B} field along the $c$ axis and in the $bc$ plane and (ii) when the {\bf B} field lies in the $ab$ plane (though not along the $a$ or $b$ axes). The first discrepancy might arise from the observation that Zr $d$ orbitals contribute to the top two valence bands by about 20\% of the total electron density according to our DFT calculations, whereas the lowest-order $k \cdot p$ model \cite{RYChen_PRL2015} was constructed based on Te $p$ orbitals only. The second discrepancy arises from the extra symmetry imposed on the lowest $k \cdot p$ model due to truncation of higher-order terms.

\begin{figure}[htb]
\begin{center}
\includegraphics[width=0.3\textwidth]{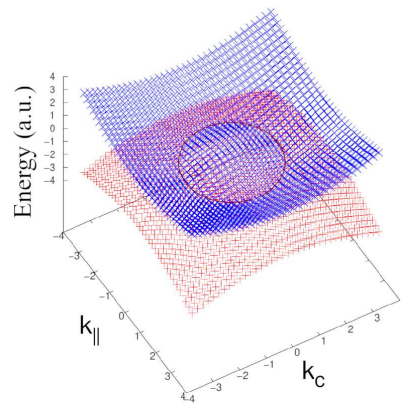}
\caption[Bab]{Calculated gapless nodal points in the $k_{\parallel}$-$k_c$ plane when the {\bf B} field is tilted by $45^{\circ}$ from the $a$ axis in the $ab$ plane, using the lowest-order $k \cdot p$ model, where $k_{\parallel}$ is the direction along the $k_a=k_b$ line. The nodal ring is marked in red. All parameter values are set to unity for simplicity.}
\label{fig:kp}
\end{center}
\end{figure}

When the {\bf B} field aligns with either the $a$ or $b$ axis, a nodal ring appears in the corresponding mirror plane. Interestingly, in the
lowest-order $k \cdot p$ model, we find that this nodal ring persists as long as the {\bf B} field lies in the $ab$ plane, even though there
is {\it no} mirror symmetry that protects the nodal ring when the {\bf B} field is away from the $a$ or $b$ axis. For example, Fig.~\ref{fig:kp}
shows the calculated nodal ring using the lowest $k \cdot p$ model when the angle between the {\bf B} field and the $a$ axis is $45^{\circ}$.
Compare this figure to the nodal structure obtained from the WFTB model [Fig.~\ref{fig:Bab}(h)] for the same {\bf B} field direction. Our study
shows that continuous $U(1)$ symmetry is present in the lowest-order $k \cdot p$ model, which rotates both the band structure and the
{\bf B} field together about the $c$ axis. This symmetry is represented by
\begin{equation}
U(\theta) {\cal H}_{\mathrm{kp}}({\bf k},{\bf  B}) U^{\dagger}(\theta) = {\cal H}_{\mathrm{kp}}(R(\theta) {\bf k},R^{-1}(\theta) {\bf B})
\end{equation}
where
\begin{equation}
U(\theta) =
\begin{bmatrix}
e^{-i\theta/2} & 0 & 0 & 0 \\
0 & e^{i\theta/2} & 0 & 0 \\
0 & 0 & e^{-i\theta/2} & 0 \\
0 & 0 & 0 & e^{i\theta/2}
\end{bmatrix},
\end{equation}
and
\begin{equation}
R(\theta) =
\begin{bmatrix}
\cos{\theta} & \sin{\theta} & 0 \\
-\sin{\theta} & \cos{\theta} & 0 \\
0 & 0 & 1
\end{bmatrix}.
\end{equation}
For simplicity, we assume an isotropic Fermi velocity and an isotropic $g$-factor, but a similar result holds in general cases.
However, higher-order terms in the $k \cdot p$ model would break the $U(1)$ symmetry, and they would gap out the nodal ring unless
there is mirror symmetry. Neither the WFTB model nor the crystal structure of ZrTe$_5$ have such $U(1)$ symmetry.

\subsection{Mirror anomaly}\label{sec5:burkov}

Recently, Burkov \cite{Burkov2018} has pointed out an additional quantum anomaly referred to as mirror anomaly inherent in Dirac semimetals with mirror symmetry, independent of their type or origin such as topological \cite{BJYang2014}, nonsymmorphic \cite{SMYoung_PRL2012}, or accidental, based on the linearized Dirac Hamiltonian. In a Dirac semimetal, the chirality operator $\gamma^5=i{\gamma^0}{\gamma^1}{\gamma^2}{\gamma^3}$ (which projects chirality of the two Weyl fermion components) \cite{Burkov2018,Burkov2017}
commutes with only one of the spin components. When the {\bf B} field rotates from this spin direction to the perpendicular
axis with which mirror symmetry is present, the Weyl points abruptly transform into a nodal ring protected by the mirror symmetry.
Furthermore, in the Dirac Hamiltonian, the positions of the Weyl nodes do not change with rotation angle until they become the nodal ring. As a consequence, intrinsic AHC was predicted to show a singular behavior as a function of the {\bf B} field direction.

In the lowest-order $k \cdot p$ model for ZrTe$_5$, Eq.~(\ref{eq:Hkp}), the spin component that commutes with the chirality operator is
the $c$ or $z$ component. Therefore, in the $k \cdot p$ model, the mirror anomaly dictates the abrupt transformation of a pair of Weyl points
into a nodal ring as well as singular intrinsic out-of-plane AHC when the {\bf B} field rotates from the $c$ axis to the $a$ or $b$ axis.
The difference between the WFTB model and the $k \cdot p$ model in this context is that the positions of the Weyl points noticeably change with the ${\bf B}$-field orientation when the field is in the $bc$ plane in the WFTB model.

%In addition, the WFTB model shows the sudden transformation of a nodal ring to another nodal ring via two Weyl points as the {\bf B} field sweeps
%from the $a$ to the $b$ axis.

\section{Intrinsic Anomalous Hall effect}\label{sec6}

When Weyl points are present at the Fermi level, the material can manifest large AHC despite the point-like Fermi surface, which serves as
an important experimental signature for Weyl semimetals. In general, broken time reversal symmetry along with a finite-volume Fermi surface
may give rise to nonzero AHC whether Weyl points are present or not.

\subsection{Numerical calculation of AHC}\label{sec6:ahc}

We numerically compute the AHC $\sigma_{ac}$ of ZrTe$_5$ under an external {\bf B} field based on our WFTB model, as a function of
chemical potential as well as the direction of {\bf B} field. In the next two subsections, we separately present our results in the cases of isotropic and anisotropic $g$ factor.
We consider $\sigma_{ac}$ because the $ac$ plane is perpendicular to the stacking direction, which is experimentally the most relevant plane \cite{RYChen_PRL2015,Liang2018}. We focus on the \emph{intrinsic} part of the AHC
\cite{Di_Xiao2010,Nagaosa2010} which depends only on the Berry curvature:
\begin{eqnarray}
\sigma_{ac} &=& \frac{e^2}{\hbar}\sum_{n=1}^{\rm{occ}} \int_{\rm{BZ}} \frac{d^3 k}{(2\pi)^3} f({\cal E}_n({\bf k})-\mu) \Omega_{n,b}({\bf k}),
\label{eq:ahc} \\
            &=& \frac{e^2}{h} \int_{k_b} \frac{dk_b}{2\pi} C(k_b) ,
\label{eq:ahc-2}
\end{eqnarray}
where $f({\cal E}_n({\bf k})-\mu)$ is the Fermi-Dirac distribution with chemical potential $\mu$, and $\Omega_{n,b}({\bf k})$ is the
$b$-component of the Berry curvature coming from the $n$-th band. Here $C(k_b)$ is the Chern number (considering all occupied bands)
calculated at a given $k_b$ plane.
We assume that temperature is zero. Regarding the 3D integral of the whole first BZ in Eq.~(\ref{eq:ahc}), we perform 2D integrals in the ${k_a}$-${k_c}$ plane ($\sigma_{ac,k_b}^{\rm{2D}}=C(k_b)\frac{e^2}{h}$) at fixed $k_b$ planes and integrate the 2D integrals along the
$k_b$ direction. In this calculation we separate the first BZ into two regions such as near the
$\Gamma$ point and away from the $\Gamma$ point, and use a finer (coarser) $k$-mesh for the former (latter) region.
%In the region near the $\Gamma$ point we use a fine mesh of up to $\Delta{k_b}$=1$\times$10$^{-5}$$\cdot$$2\pi \cdot \AA^{-1}$ and $\Delta{k_a},\Delta{k_c}$=1.25$\times$$10^{-5}$$\cdot$$2\pi \cdot \AA^{-1}$, whereas outside this
%region we use a mesh of $\Delta{k_b}$=0.001$\cdot$$2\pi \cdot \AA^{-1}$ and $\Delta{k_a,k_c}$=0.0005$\cdot$$2\pi \cdot \AA^{-1}$.
%Considering the locations of the gapless nodal points, we have a fine enough $k$-point mesh.

When the Fermi surface is composed of a pair of isolated Weyl points with topological charge $\pm \chi$, $|\sigma_{ac}|$ is known to be
$\frac{e^2}{h} \frac{2k_b^{\rm{WP}}}{2\pi} |\chi|$, where $2k_b^{\rm{WP}}$ is the separation between the two Weyl points projected
onto the $b$ axis. This is because each 2D plane parallel to the $ac$ plane which lies between the two Weyl points contributes Chern number of $\chi$ or $-\chi$ to the integral in Eq.~(\ref{eq:ahc-2}), whereas the planes which do not lie between the two Weyl points have zero Chern number. In this case, the intrinsic AHC is simply proportional to the separation between the Weyl points along the $b$ axis. However, if the Fermi surface has a nonzero volume, there is no such a simple expression for $\sigma_{ac}$ and the AHC must be numerically computed.

Equation~(\ref{eq:ahc}) enforces that $\sigma_{ac}$ is strictly zero if the {\bf B} field is parallel to the $ac$ plane due to the
$C_{2b} {\cal T}$ symmetry. The $C_{2b} {\cal T}$ symmetry maps the Berry curvature component $\Omega_{b}(k_a,k_b,k_c)$ into $-\Omega_{b}(k_a,-k_b,k_c)$, making the integral in
Eq.~(\ref{eq:ahc}) vanish. Therefore, the nonzero AHC observed in Ref.~\cite{Liang2018} for the in-plane {\bf B} fields
(i.e. parallel to the $ac$ plane) must originate from the non-intrinsic part of the AHC and/or nonlinear effects.

\subsection{Calculated AHC in the case of isotropic $g$ factor}

\begin{figure*}[htb]
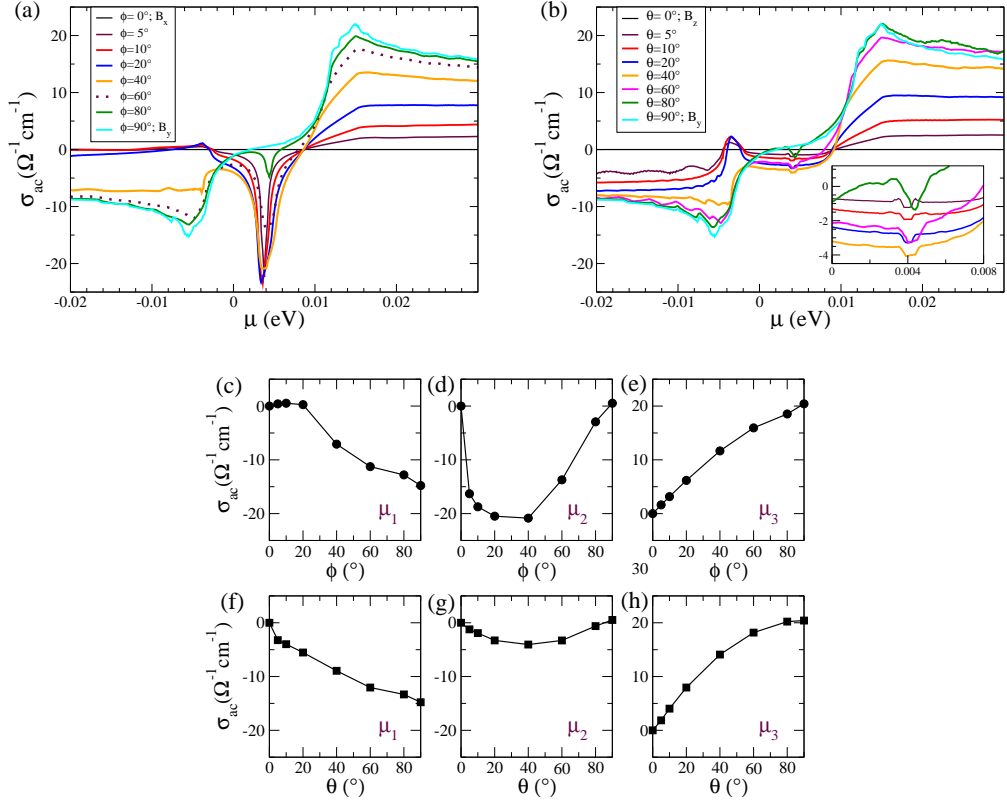

\begin{center}
%\vspace{0.2truecm}
\includegraphics[width=0.35 \textwidth]{Fig9a_new.eps}
\hspace{0.5truecm}
\includegraphics[width=0.35 \textwidth]{Fig9b_new.eps}

\vspace{0.5truecm}

\includegraphics[width=0.45\textwidth]{Fig9c_new.eps}
\caption{WFTB-calculated AHC $\sigma_{ac}$ as a function of chemical potential $\mu$ and tilting angles $\phi$ and $\theta$ when the {\bf B} field is parallel to (a) the $ab$ plane or (b) the $bc$ plane, in the case of isotropic $g$ factor.
Here $\phi$ is the angle between the {\bf B} field and the $a$ axis, and $\theta$ is the angle
between the {\bf B} field and the $c$ axis. The Fermi level is set to $\mu=0$. (c)-(h) Calculated $\sigma_{ac}$ vs $\phi$ and $\theta$ at three chemical potential values, $\mu_1$ ($-$6 meV), $\mu_2$ (4 meV), and $\mu_3$ (14 meV). The $\mu_2$ value is close to the type-I Weyl point energy.
The inset in (b) shows zoom-in near the $\mu_2$ value.}
\label{fig:AHC-1}
\end{center}
\end{figure*}

\begin{figure}[htb]
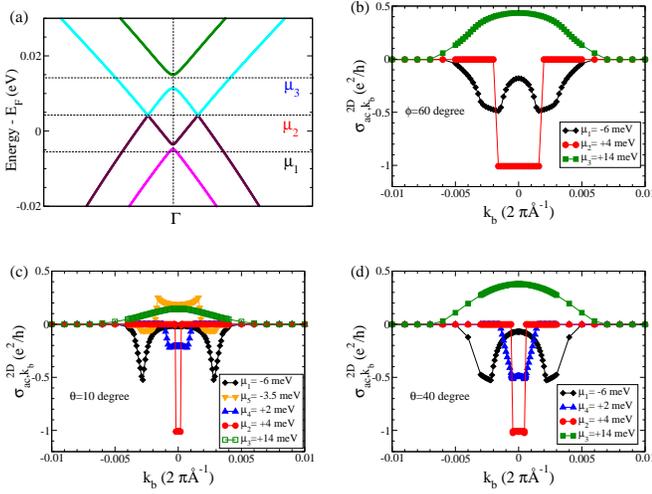

\begin{center}
\includegraphics[width=0.23 \textwidth]{Fig10a.eps}
\hspace{0.2truecm}
\includegraphics[width=0.23 \textwidth]{Fig10b.eps}

\vspace{0.5truecm}

\includegraphics[width=0.23 \textwidth]{Fig10c.eps}
\hspace{0.2truecm}
\includegraphics[width=0.23 \textwidth]{Fig10d.eps}
\caption{(a) WFTB-calculated band structure along the Weyl-point separation direction and (b) $\sigma_{ac,k_b}^{2D}$ at fixed $k_b$ values
vs $k_b$ for three different chemical potential values, when the {\bf B} field is tilted from the $a$ axis by 60$^{\circ}$ in the $ab$
plane. (c)-(d) Calculated $\sigma_{ac,k_b}^{2D}$ vs $k_b$ for different chemical potential values, when the
{\bf B} field is tilted from the $c$ axis by 10$^{\circ}$ or 40$^{\circ}$ in the $bc$ plane, respectively. Here $\mu_5=-3.5$~meV
corresponds to the energy of the small positive AHC peak right below the Fermi level for $\theta$=5, 10, and 20$^{\circ}$. The
type-II Weyl-point energy is $-$4.5~meV. In (a)-(d) we consider the isotropic $g$ factor.}
\label{fig:AHC-2}
\end{center}
\end{figure}

Figure~\ref{fig:AHC-1}(a) shows numerically calculated $\sigma_{ac}$ as a function of chemical potential when the {\bf B} field
rotates in the $ab$ plane, in the case of isotropic $g$ factor. First of all, we find that intrinsic $\sigma_{ac}$ becomes strictly zero
independent of the chemical potential value when the {\bf B} field exactly aligns with the $a$ axis due to the symmetry argument provided
earlier. Let us first discuss features at chemical potential $\mu=-4$~meV (referred to as $\mu_2$) which coincides with the type-I Weyl point
energy. See Fig.~\ref{fig:AHC-2}(a). At this chemical potential, a sharp, prominent (negative) peak appears for all angles except for
0$^{\circ}$ and 90$^{\circ}$. The peak height varies with $\phi$, as shown in Fig.~\ref{fig:AHC-1}(d). In order to provide more insight,
we also calculate $\sigma_{ac,k_b}^{\rm{2D}}$ at different $k_b$ planes, finding that they are quantized as either 0 or $-$1 in units
of $\frac{e^2}{h}$. This result explains both the peak height $\sim \frac{e^2}{h} \frac{2k_b^{\rm{WP}}}{2\pi}$
and the angular evolution of the peak height.
The abrupt increase in the peak height is attributed to the large separation of the Weyl point along the $b$ axis as the angle increases
from zero. The peak height, however, goes to zero smoothly as the angle approaches 90$^{\circ}$ because of the smooth changes of the
Weyl point separation. See Fig.~\ref{fig:Bab}(i).
In addition to the sharp peak, smoothly rising negative and positive peaks appear near +6 meV ($\mu_1$) and +14 meV ($\mu_3$) for large
angles ($\ge$~60$^{\circ}$). See Figs.~\ref{fig:AHC-1}(a), (c), and (e). Note that there are no other Weyl nodes beyond the pair discussed
earlier. Now the $\sigma_{ac,k_b}^{\rm{2D}}$ values for $\mu_1$ and $\mu_3$ are not quantized in units of
$\frac{e^2}{h}$ [Fig.~\ref{fig:AHC-2}(b)]. However, they contribute to the $\sigma_{ac}$ value, Eq.~(\ref{eq:ahc}), via avoided level crossings as
shown in Fig.~\ref{fig:AHC-2}(a). Our result is not unusual. For example, in bcc Fe, Co, and Ni, a very large Berry curvature was found in
regions where avoided level crossings occur, and it resulted in large AHC \cite{XWang2007}.

Now when the {\bf B} field rotates from the $c$ axis in the $bc$ plane, overall features of $\sigma_{ac}$ [Figs.~\ref{fig:AHC-1}(b), (f)-(h)]
are similar to those for the above case, though with some differences. Let us first discuss $\sigma_{ac}$ at chemical potential $\mu_2$.
At this chemical potential, the height of the negative peak can be up to an order of magnitude smaller than that for the above case, as shown
in the inset of Fig.~\ref{fig:AHC-1}(b) and Fig.~\ref{fig:AHC-1}(g). The $\sigma_{ac,k_b}^{\rm{2D}}$ values are quantized as either 0 or $-$1
in units of $\frac{e^2}{h}$. For example, Fig.~\ref{fig:AHC-2}(c) and (d) show such quantization for $\theta=10$ and 40$^{\circ}$. The observed peak
height of $\frac{e^2}{h} \frac{2k_b^{\rm{WP}}}{2\pi}$ also corroborates the contributions of the type-I Weyl points which evolve with the angle as
illustrated in Fig.~\ref{fig:Bbc}(a). Next, for $0 < \mu < 8$ meV, the flat $\sigma_{ac}$ region arises from contributions of small nonzero
$\sigma_{ac,k_b}^{\rm{2D}}$ values near $\Gamma$. See upward-triangles in Fig.~\ref{fig:AHC-2}(c) and (d). Last, at $\mu=-3.5$~meV a small
positive peak appears only for small angles ($<$ 40$^{\circ}$). This peak is associated with the type-II Weyl nodes arising
from the crossings of the two valence bands. See Fig.~\ref{fig:Bbc}(d). At higher angles, the type-II Weyl points with opposite chirality
are annihilated, and the $\sigma_{ac}$ peak vanishes accordingly.

\subsection{Calculated AHC in the case of anisotropic $g$ factor}

\begin{figure*}[htb]
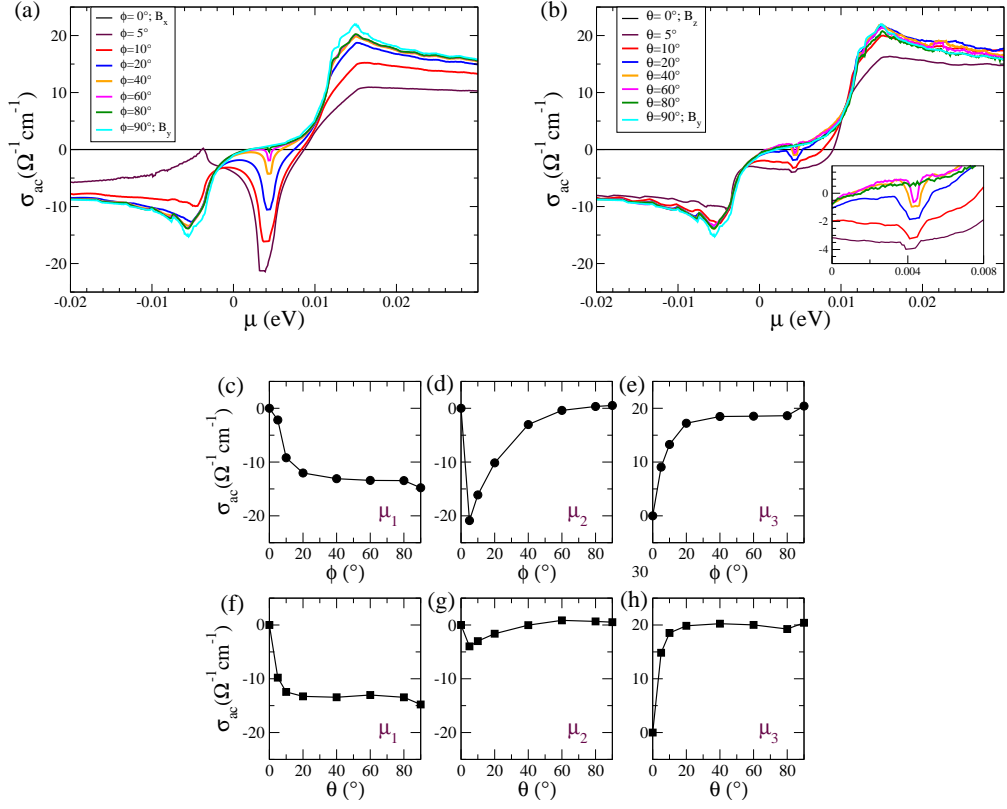

\begin{center}
\includegraphics[width=0.35 \textwidth]{Fig11a_new.eps}
\hspace{0.5truecm}
\includegraphics[width=0.35 \textwidth]{Fig11b_new.eps}

\vspace{0.5truecm}

\includegraphics[width=0.45\textwidth]{Fig11c_new.eps}
\caption{WFTB-calculated AHC $\sigma_{ac}$ as a function of $\mu$, $\phi$ and $\theta$ when the {\bf B} field is parallel to the $ab$ plane (a)
or the $bc$ plane (b), in the case of {\it anisotropic} $g$ factor. (c)-(h) Calculated $\sigma_{ac}$ vs $\phi$ and $\theta$ at
$\mu_1$ ($-$6 meV), $\mu_2$ (4 meV), and $\mu_3$ (14 meV). The $\mu_2$ value is close to the type-I Weyl point energy.
The inset in (b) shows zoom-in near the $\mu_2$ value.}
\label{fig:AHC-3}
\end{center}
\end{figure*}

% Sep 26 2019: Anisotropic g factor case. Emphasize distinctive points. A remark regarding the comparison to experiment.
% maybe we mention the experimental data for resistivity and theory conductivity in introduction?? not sure. Maybe here.
% Definitely include the difficulty with comparison to exp data in conclusion.

Figure~\ref{fig:AHC-3}(a) shows our calculated $\sigma_{ac}$ as a function of $\mu$ and $\phi$, using the following anisotropic $g$ factors:
$g_x$=3.19 \cite{YLiu2016}, $g_y$=21.3 \cite{RYChen_PRL2015,YLiu2016}, and $g_z$=2.0.
In this case, the Zeeman energy remains to be fixed as 10~meV. In other words,
$\sqrt{g_x^2 B_x^2 + g_y^2 B_y^2 + g_z^2 B_z^2}\frac{\mu_B}{2}$=10~meV. When the {\bf B} field rotates from the $a$ axis in the $ab$ plane,
the height of the sharp AHC peak at $\mu_2$ abruptly increases with the angle and then it immediately starts to decrease. See
Fig.~\ref{fig:AHC-3}(d). On the other hand, the height of the smooth peaks at $\mu_1$ and $\mu_3$ sharply increases with the angle and
then it saturates at a small angle. See Fig.~\ref{fig:AHC-3}(c) and (e). These features can be explained using the Weyl-point positions
and avoided level crossings, similarly to the case of isotropic $g$ factor.

A similar trend appears when the {\bf B} field is tilted in the $bc$ plane [Figs.~\ref{fig:AHC-3}(b), (f)-(h)]. One small difference is
shown in the inset of Fig.~\ref{fig:AHC-3}(b). The small AHC peak at $\mu_2$ vanishes for very large angles such as
80~$\le \theta <$~90$^{\circ}$, because the $b$ component of the Weyl point position becomes zero.

Overall the angular dependence of $\sigma_{ac}$ with the anisotropic $g$ factor qualitatively differs from that with the isotropic $g$ factor in both $ab$ and $bc$ planes. This is attributed to a much larger contribution of the $b$ component of {\bf B} field for a given angle. For the same reason, with the anisotropic $g$ factor, the type-II Weyl nodes are not formed except for extremely small angles away from the $c$ axis. Note that the type-II Weyl nodes with opposite chirality are annihilated at $\theta \ge$~40$^{\circ}$ in the case of isotropic $g$ factor.

\section{Conclusion}\label{sec7}

In summary, we develop a WFTB model for 3D ZrTe$_5$ from first-principles calculations, considering both Zr $d$ and Te $p$ orbitals. Based
on the WFTB model, we investigate Zeeman-splitting induced topological phases and the evolution of the topological nodal structures as a
function of the orientation of {\bf B} field (beyond the crystal axes). We find an abrupt transformation of a nodal ring to a pair of
Weyl nodes as the {\bf B} field is rotated from either the crystal $a$ or $b$ axis. At some {\bf B} field directions type-II nodal structures
are identified from crossings of the valence bands. Comparing the calculated topological phases with those from the linearized
$k \cdot p$ model, we find that the latter model does {\it not} capture the correct topological phases when the {\bf B} field is
rotated in the $ab$ or in $bc$ plane. We also numerically compute the intrinsic part of the AHC, $\sigma_{ac}$, as a function of chemical
potential, when the {\bf B} field is tilted within the $ab$ or $bc$ plane. The calculated results can be compared to the experimental data
when the experimental anomalous Hall resistivity $\rho_{ac}^{\rm{AHE}}$ is properly converted into $\sigma_{ac}$, which requires
the knowledge of longitudinal resistivity. Our findings may also provide insight into Zeeman-splitting-induced topological phases and their
consequences in other Dirac semimetals with mirror symmetries.

\begin{acknowledgments}
Y.C. was supported by the Virginia Tech ICTAS Fellowship. The computational support was provided by San Diego Supercomputer Center (SDSC) under DMR060009N and VT Advanced Research Computing (ARC).
\end{acknowledgments}

%\begin{comment}
%\end{comment}
%\bibliography{references4_JV}
%\bibliographystyle{ieeetr}

%

\end{document}